\documentclass[useAMS,usenatbib]{mn2e}
\usepackage{graphicx}
\usepackage{amsmath}
\usepackage{txfonts}
\usepackage{epsf, epsfig}
\usepackage{color}

\usepackage{color}
\usepackage{enumitem}
\usepackage[]{hyperref}
\usepackage{verbatim}
\usepackage{amsmath,amstext,mathrsfs}
\usepackage{afterpage}   
\usepackage{marginnote}
\usepackage{makecell}
\usepackage{subfigure}

\title[Tracing Magnetic Field with Synchrotron Polarization Gradients]{Tracing Magnetic Field with Synchrotron Polarization Gradients: Parameter Study}
\author[Zhang et al. ]{Jian-Fu Zhang$^{1,2}$, Alex Lazarian$^{2}$, Ka Wai Ho$^{3}$, Ka Ho Yuen$^{2}$, Bo Yang$^2$, Yue Hu$^{2,4}$\\
$^1$Department of Physics, Xiangtan University, Xiangtan, Hunan 411105, China; jfzhang@xtu.edu.cn\\
$^2$Astronomy Department, University of Wisconsin, Madison, WI 53711, USA; alazarian@facstaff.wisc.edu\\
$^3$Department of Physics, Chinese University of Hong Kong, Hong Kong, China;\\
$^4$Physics Department, University of Wisconsin, Madison, WI 53711, USA}

\date{Accepted xxx. Received xxx; in original form xxx}

\begin{document}
\pagerange{\pageref{firstpage}--\pageref{lastpage}} \pubyear{2019}

\maketitle

\begin{abstract}
We employ synthetic observations obtained with MHD simulations to study how to trace the distribution of turbulent magnetic fields using the synchrotron polarization gradient techniques suggested by Lazarian \& Yuen (2018). Both synchrotron polarization gradients and its derivative gradients with regard to the squared wavelength $\lambda^2$ are used to explore the probing ability of the technique in magnetic fields from sub- to super-Alfv{\'e}nic turbulence. We focus on studies that involve multi-frequency measurements in the presence of strong Faraday rotation and show the ways of how to recover the projected mean magnetic fields in the plane of the sky and the local magnetic fields within a tomographic slice. We conclude that the new techniques can successfully reconstruct the 3D magnetic field within our Milky Way and other galaxies. This paper opens an avenue for applying our new techniques to a large number of data cubes such as those from the Low Frequency Array for Radio astronomy and the Square Kilometer Array. 
\end{abstract}


\section{Introduction}
As is well known, magnetohydrodynamic (MHD) turbulence occurs naturally in most of astrophysical flows due to a large spatial scale resulting in a very high Reynolds number (e.g., \citealt{Cho2003LNP,Lazarian15}). The presence of magnetic turbulence has been influencing and even changing many key astrophysical processes, such as star formation (\citealt{Elmegreen04,Mckee07,Lazarian12}), heat transfer (\citealt{Narayan01,Lazarian06}), propagation and acceleration of cosmic rays (\citealt{Schlickeiser02,Yan02}), as well as turbulent magnetic reconnection (\citealt{LazarianV99,Kowal09,Lazarian15} for a review). Therefore, studying MHD turbulence theory and its implications to astrophysics is an interesting and practical applicable topic. 

Synchrotron radiation fluctuation carrying the statistical information of MHD turbulence would be emitted with the movement of relativistic electrons in a turbulent magnetic field. The linearly polarized synchrotron emission can be characterized in terms of the synchrotron intensity (Stokes parameter $I$) and polarization intensity ($P=\sqrt{Q^2+U^2}$ by Stokes parameters $Q$ and $U$). The synchrotron polarization together with Faraday rotation can be used to explore the properties of magnetic field, which is termed the traditional Faraday rotation synthesis (e.g., \citealt{Burn66,Brentjens05,Frick11,Beck12}). This technique recently applying to polarized observational data provides a significant insight into magnetic field structures of galaxies (e.g., \citealt{Fletcher11,Beck13,Haverkorn15} for a review) and the properties of the interstellar medium (ISM, e.g., \citealt{Jelic15,VanEck17,Dickey18}).

Different from the traditional Faraday rotation technique, the statistics of the synchrotron intensity (\citealt{LP12}, hereafter LP12) and polarized intensity (\citealt{LP16}, hereafter LP16) are formulated  in the real and wavelength spaces providing a quantitative analytical description of synchrotron fluctuation. The corresponding theoretical predictions regarding the synchrotron intensity are successfully tested in \cite{Herron16}. Furthermore, it was demonstrated that numerical results are in good agreement with theoretical predictions of LP16 (\citealt{Lee16,Zhang16,ZhangL18}), in which the polarization frequency analysis (PFA) and polarization spatial analysis (PSA) techniques can be used to extract the properties of MHD turbulence, such as the spectral slope and correlation scale. The new statistical techniques suggested in LP16 have been applied to the ISM (\citealt{XuZhang16}) and depolarization of blazars (\citealt{Guo17}).

In addition, statistical techniques for the analysis of polarization gradients, i.e., the gradient of complex polarized vector has been applied to determine the sonic Mach number of the interstellar turbulence (\citealt{Gaensler11,Burkhart12,Iacobelli14,Sun14,Robitaille15,Herron17}). Recently, more sophisticated constructions of Stokes parameters were discussed in \cite{Herron18a,Herron18b} as possible means of exploring magnetized ISM. 

A significant breakthrough took place in the field recently in \cite{Lazarian17} and Lazarian \& Yuen (2018a, henceforth LY18) identifying the relation between the direction of the Synchrotron Intensity Gradients (SIGs), Synchrotron Polarization Gradients (SPGs) as well as Synchrotron Polarization Derivative Gradients (SPDGs) and the underlying magnetic field. In particular, it was shown that in turbulent magnetized media all these gradients are directed perpendicular to the magnetic field. As the direction of these gradients is not affected by the Faraday rotation, this opened a way to study magnetic field without accounting for the Faraday rotation. Therefore, uniquely, the gradient measurements of synchrotron intensity or synchrotron at a {\it single frequency} can establish the magnetic field direction in the emitting synchrotron region, irrespectively, of how strong the Faraday rotation is there. Moreover, by employing the effect of Faraday depolarization, LY18 discussed a way of obtaining 3D distribution of the magnetic field in the emitting volume. 

The prospects of the SPGs and SPDGs for studying magnetic fields in the ISM motivate us to provide a detailed numerical study of the properties of the gradients for a variety of turbulence setups.
In what follows we explore the statistics of gradients for a variety of sonic and Alfv{\'e}n Mach numbers.

The content of this paper is outlined as follows. In the next section, we provide a brief method description including MHD turbulence theoretical fundamental and gradients of magnetic intensity, synchrotron polarization and its derivative gradients, as well as gradient measurement technique. The numerical results are presented in Section 3. Sections 4 and 5 are the discussion and summary, respectively. 

\section{Method Descriptions}
\subsection{MHD turbulence and gradients of magnetic field strength} \label{MHDGMI}
The modern MHD turbulence theory present a collection of anisotropic eddies (\citealt{Gold95}, hereafter GS95), which are aligned with the direction of magnetic fields. For the gradient techniques that we discuss in the paper it is important that the eddies are aligned with the local magnetic field surrounding the eddies. The latter element is far from trivial as all the theories of turbulence before GS95, as well as GS95 theory itself, implicitly assume that the alignment is in terms of the mean magnetic field. At the same time, this {\it local} alignment follows directly from the theory of turbulent magnetic reconnection  (\citealt{LazarianV99}, henceforth LV99). Indeed, LV99 theory predicts that the reconnection in turbulent fluid takes place over just one eddy turnover time. As a result, magnetic field does not prevent to the motion of eddies that mix up magnetic field lines perpendicular to the direction of the magnetic field that passes through the eddy. In the situation of random driving the eddies that experience less resistance obtain more energy. Therefore, most of the energy of Alfv{\'e}nic motion is concentrated in such eddies that are perpendicular to the local magnetic field. This alignment of the eddies in respect to the local magnetic field is confirmed by numerical simulations (\citealt{ChoV00,Maron01,ChoLV02}. The consequence of the aforementioned alignment is that the gradients of magnetic field strength that, for the sake of simplicity, we shall call simply "B-gradients", are aligned perpendicular to the magnetic field direction. Therefore, in the synchrotron gradient techniques that we discuss in this paper we turn the magnetic field gradients 90 degrees to identify the magnetic field direction.  

Naturally, magnetic field is only important for determining the direction of eddies only when the energy of magnetic field over the volume of the eddy is larger or comparable with the kinetic energy of the eddy. The measure that reflects this is the Alfv{\'e}nic Mach number, $M_{\rm A}=V_{\rm L}/V_{\rm A}$ , where $V_{\rm L}$ is an injection velocity of turbulence driving at the scale $L_{\rm inj}$ and $V_{\rm A}= B/\sqrt{4\pi \rho}$ is an Alfv{\'e}n velocity relevant to the magnetic field $B$ and the plasma density $\rho$. We could describe different regimes of MHD turbulence by using this parameter. The GS95 theory considers incompressible turbulence with $M_{\rm A}=1$ while its generalizations for  $M_{\rm A}<1$ and $M_{\rm A}>1$ can be found in \cite{LazarianV99} and \cite{Lazarian06}. The applicability of the model of realistic compressible turbulence were obtained in \cite{Lithwick01} and \cite{Cho2002PRL,ChoL03}. The latter two papers demonstrated that while density is seriously modified by compressibility, the magnetic and velocity fluctuations of Alfv{\'e}n and slow modes are only marginally different from the incompressible case.\footnote{Fast modes have different statistics, but numerical simulations show that the amount of energy in fast modes is less than in Alfv{\'e}n and slow modes (\citealt{Cho2002PRL,Kowal10}). } In terms of gradient technique in this paper, this gives us confidence in applying the technique to the realistic compressible ISM. 

The $M_{\rm A}<1$ case, i.e. the case of turbulence driven with sub-Alfv{\'e}nic velocities at injection scale $L_{\rm inj}$ has a range of {\it weak} turbulence that spans from $L_{\rm inj}$ to the transition scale $l_{\rm trans}=L_{\rm inj}M_{\rm A}^2$. Over this range the perturbation of magnetic field are quasi-2D and they are perpendicular to magnetic field (LV99, \citealt{Galtier00}). Thus the gradients of magnetic field are also aligned perpendicular to the magnetic field. For scales less than $l_{\rm trans}$ the strong sub-Alfv{\'e}nic turbulence is present. The eddies of this turbulence are more elongated along the magnetic field compared to the original GS95 picture that corresponds to $M_{\rm A}=1$. Indeed LV99 showed that over the inertial range of $[l_{\rm trans}, l_{\rm diss}]$, where, $l_{\rm diss}$ is the turbulence dissipation scale. The relation between the scale of the eddy extend along the magnetic field $l_\|$ and its transversal extend $l_\bot$ is  
\begin{equation}
\label{anis}
l_{\|}\approx L_{\rm inj}\left(\frac{{ l_\perp}}{L_{\rm inj}}\right)^{2/3} M_{\rm A}^{-4/3},
\end{equation}
when $M_A=1$ returns to the original GS95 relation. The turbulent velocity of eddies at scales less than $l_{\rm trans}$ obeys the following relation:
\begin{equation}
{v_\perp}=V_{\rm A} \left(\frac{{ l_\perp}}{L_{\rm inj}}\right)^{1/3} M_{\rm A}^{4/3}=V_{\rm L} \left(\frac{{ l_\perp}}{L_{\rm inj}}\right)^{1/3} M_{\rm A}^{1/3},
\label{vel_strong}
\end{equation}
which demonstrates Kolmogorov-type ($v_l\propto l_{\perp}^{1/3}$) cascade perpendicular to local magnetic field. This is an important scaling that is being employed in the gradient technique. Indeed, the B-gradients scale as $l_\bot^{1/3}/l_{\bot}\sim l_\bot^{-2/3}$ which shows that the B-gradients arising at the smallest scales, i.e. telescope resolution scales, dominate the signal.  

The opposite case corresponding to super-Alfv{\'e}nic turbulence $V_{\rm L}> {V}_{{\rm{A}}}$, i.e., ${M}_{{\rm{A}}}>1$. For a limiting case of ${M}_{{\rm{A}}}\gg 1$, since the magnetic field is weak and has a marginal influence on MHD turbulence, the turbulence at scales close to the injection scale has an essentially hydrodynamic Kolmogorov property, i.e., $v_l=V_{\rm L}(l/L_{\rm inj})^{1/3}$. The hydrodynamic characteristic of the turbulence cascade changes at the scale
\begin{equation}
\label{SupLA}
l_{\rm A}=L_{\rm inj}M_{\rm A}^{-3},
\end{equation}
where the turbulent velocity is equal to the Alfv{\'e}n velocity, $v_l=V_{\rm A}$ (\citealt{Lazarian06}). Naturally, for hydrodynamic-like turbulent motions at scales larger than $l_{\rm A}$, the directions of gradients of magnetic field are not correlated with magnetic field direction. Indeed, when hydrodynamic motions dominate, no alignment of eddies and magnetic field is expected. This situation changes  at scales less than $l_{\rm A}$. At such scales the magnetic field becomes again important and one can expect to see the alignment of the B-gradients perpendicular to magnetic field. Therefore, if one filters out large scale motions, e.g. by using spatial filtering, it is potentially possible to trace magnetic field with the B-gradients. This possibility was demonstrated in \cite{Lazarian17}, but in practical numerical testing this presents a challenge for sufficiently large $M_{\rm A}$. Indeed, in most cases the resolution of simulations is such that $l_{\rm A}$ becomes too close to $l_{\rm diss}$ or even less than $l_{\rm diss}$. In the latter case the turbulence is similar to the hydrodynamic one up to the dissipation scale and no magnetic field tracing is possible in simulations. In realistic astrophysical settings $l_{\rm diss}$ is usually much smaller than $l_{\rm A}$, however. Thus we expect that B-gradients can trace magnetic field in the case of $M_{\rm A}>1$ without any difficulty. However, testing this is numerically very challenging. 

\subsection{Synchrotron polarization and its derivative gradients}\label{SynTheor}
The fluctuation of polarized synchrotron radiation reflects the information of magnetic turbulence fluctuation. In this work, we consider that non-thermal relativistic electrons have a power-law energy distribution of $N_e\propto\gamma^{-p}$, where $p$ is a spectral index of electrons, and adopt the formulae commonly used in the study of synchrotron polarization emission (see the Appendix of \citealt{Waelkens09} and LP16). Based on Stokes parameters $Q$ and $U$ to describe a linear polarization, we consider the synchrotron polarization intensity vector in the form of $\textbf{\textit{P}}=Q+iU$. Following LP16, we have 
\begin{equation}
\textbf{\textit{P}}(\textbf{\textit{X}},\lambda^2)=\int^{L}_{0}dzP_i(\textbf{\textit{X}},z){e}^{2i\lambda^2\varPhi(\textbf{\textit{X}},z)}, \label{PEq}
\end{equation}
where $L$ is the extended scale along the line of sight ($z$-axis direction). $P_i(\textbf{\textit{X}},z)$ stands for the intrinsic polarized intensity density treated as wavelength-independent; this simplified treatment would not change our statistical results (\citealt{ZhangL18}). Therefore, the wavelength dependence explored in this work is only associated with Faraday depolarization processes, defining Faraday rotation measure as 
\begin{equation}
\varPhi(\textbf{\textit{X}},z)={e^3\over2\pi m_{\rm e}^2c^4}\int^z_{0}dz'n_{\rm e}(\textbf{\textit{X}},z')B_z(\textbf{\textit{X}},z'), \label{RM}
\end{equation}
where $n_{\rm e}$ and $B_{ z}$ are the thermal electron density and the parallel component of magnetic field, respectively.  Using equations (\ref{PEq}) and (\ref{RM}), and combining with synchrotron emissivity formulae provided in LP16, we can obtain the (integrated) synchrotron polarization intensity $P=\sqrt{Q^2+U^2}$ along the line of sight, which has an advantage of phase independence to recover magnetic field distribution in turbulent media. 

LP16 introduced a one-radian definition as the condition for de-correlation of Faraday rotation 
\begin{equation}
\label{eq:oneradcon}
\lambda^2\varPhi = 0.81\lambda^2 \int_0^{L_{\rm eff}} dz n_{\rm e} B_z\approx 1.
\end{equation}
According to equation (\ref{eq:oneradcon}), the ratio of the scale that is sampled by polarization to the extent of the emitting region can be written as
\begin{equation}
\label{eq:el}
\frac{L_{\rm eff}}{L} \sim \frac{1}{\lambda^2L} \frac{1}{\phi},
\end{equation}
where, $\phi= {\rm max}(\sqrt{2} \sigma_\phi,\bar{\phi})$, $\sigma_\phi$ is the dispersion of random magnetic field and $\bar{\phi}$ is the mean Faraday rotation measure density. The strong and weak Faraday rotation can be characterized by $L_{\rm eff}/L<1$  and $L_{\rm eff}/L>1$, respectively. On the basis of equation (\ref{eq:el}), if providing Stokes parameters $Q$ and $U$ at
 two different frequencies, we can trace information of magnetic field fluctuation originating from different line-of-sight depths. In general, using the differences of an observable at two near frequencies will therefore provide a measure of local mean magnetic field in a slice (see LY18 for more details).
 
When the condition of $\Delta L / L \sim \Delta (\lambda^2\phi) / (\lambda^2\phi) \ll 1$ is satisfied, synchrotron polarization intensities for a chosen wavelength $\lambda$ with the Faraday depolarization effect could be written as 
\begin{equation}
\begin{aligned}
\label{eq:leffP}
\textbf{\textit{P}}(\lambda) = \int_0^{L_{\rm eff}(\lambda)} dz P_i(\textbf{\textit{X}}, z) e^{2i\lambda^2 \varPhi(X, z)}.
\end{aligned}
\end{equation}
By using equation (\ref{eq:leffP}), the spatial gradient of polarization intensity, $\nabla P$, provides a measure of the cumulative contribution of $P_i$ plus an extra Faraday rotation fluctuation. In the case of a strong Faraday rotation, $L_{\rm eff}/L<1$,  equation (\ref{PEq}) can be split two parts, in which the only part of $z<L_{\rm eff}$ suffers from strong Faraday depolarization while the part of $z>L_{\rm eff}$ does not contribute to the average of the polarization intensity map (see Figs. 1 and 15 in LY18 for an intuitive illustration). Therefore, the difference of $\textbf{\textit{P}}(\lambda_1)$ and $\textbf{\textit{P}}(\lambda_2)$ can be expressed by  
\begin{equation}
\begin{aligned}
\label{eq:diffP}
\Delta \textbf{\textit{P}} 
&\approx \int_{L_2}^{L_1} dz P_i(\textbf{\textit{X}}, z) e^{2i\lambda^2\Phi(\textbf{\textit{X}}, z)}.
\end{aligned}
\end{equation}
The local mean magnetic field between $z\in [L_2,L_1]$ projected in the plane of the sky is traced by 
\begin{equation}
\label{eq:SPDGa}
\nabla \frac{d|P|}{d\lambda^2} \sim \lambda^{-2} \nabla |\textbf{\textit{P}}(\lambda_1)-\textbf{\textit{P}}(\lambda_2)|,
\end{equation}
which is so-called SPDGs. In our numerical practice, we use 
\begin{equation}
\label{eq:SPDG}
\nabla \frac{|\Delta P|}{\Delta\lambda^2} =\nabla\left[\sqrt{ (\Delta Q/ \Delta \lambda^2)^2+ (\Delta U/ \Delta \lambda^2)^2}\right]
\end{equation}
to calculate the SPDGs, which would determine the direction of local mean magnetic field. 

The polarization spatial analysis of $\frac{d P(\textbf{\textit{X}})}{d \lambda^2}$ suggested in LP16 can extract fluctuation information of Faraday rotation, which is successfully tested in \cite{Lee16} and \cite{ZhangL18} by synthetic observations. In other words, two-point statistics of  $\frac{d P(\textbf{\textit{X}})}{d \lambda^2}$ presents a power-law scaling that recovers the correlation index of the Faraday rotation measure. Similarly, we can suggest the Faraday rotation gradient measure $\nabla \varPhi$ on the basis of equation (\ref{RM}), which is a limiting case of the SPDGs for long wavelengths.
However, $\varPhi$ is not a directly observable quantity, thus one can adopt the so called polarization-weighted Faraday rotation measure
\begin{equation}
\label{eq:faradw}
\tilde{\varPhi}= -i\frac{d{\rm log}[\textbf{\textit{P}}]}{d\lambda^2} = \frac{\int^{L}_0 dz \varPhi(\textbf{\textit{X}}, z) P_i(\textbf{\textit{X}}, z) e^{i\lambda^2 \varPhi(\textbf{\textit{X}}, z)}}{\int^{L}_0 dz P_i(\textbf{\textit{X}}, z) e^{i\lambda^2 \varPhi(\textbf{\textit{X}}, z)}},
\end{equation}
which is equivalent to the log-derivative of equation (\ref{PEq}). If considering the gradient of the modulus of the weighted measure, $\nabla |\tilde{\varPhi}|$, it can be found that this measure is similar to the one presented in equation (\ref{eq:SPDGa}), that is, the SPDGs can deliver Faraday rotation fluctuation in the case of the coincident synchrotron and Faraday rotation regions. 

\subsection{Gradient measure techniques}

We follow the way the synchrotron polarization gradients were calculated in LY18. They used the recipe of sub-block averaging first suggested for velocity gradients (see \citealt{YL17} for the details) to calculate the gradients of synchrotron polarization and its derivative. This technique is based on the fitting a Gaussian into the distribution of gradients calculated within a sub-block of maps. The error of fitting gives the error associated with the magnetic field tracing with the technique. For comparing the gradient directions with that of magnetic field, the gradients are turned  $90^\circ$.  

The alignment measure (AM) is introduced as (see \citealt{GL17}) 
\begin{equation}
AM=2\langle\cos^2\theta\rangle-1 \label{AMs}
\end{equation}
to determine a relative alignment between the rotated $90^\circ$ polarization gradients and magnetic field directions, which is in analogy with the method of grain alignment studies in \cite{Lazarian07}. In equation (\ref{AMs}), $\theta$ is an angle subtended by any magnetic field and the gradient directions. It is evident that $AM$ is in a range of $[-1,1]$. When $AM=-1$, the rotated gradient direction is perpendicular to the magnetic field. When $AM=1$, they have a perfect alignment, whereas random orientations result in $AM\sim0$ . 

\section{Simulation results }
Simulation 3D data cubes we used are generated from two open source ZEUS and ATHENA  codes (\citealt{Hayes06, Stone08}). They are grid-based codes for simulation of astrophysical MHD processes. By assuming a periodic boundary condition and a solenoidal injection of the turbulence driving, we set up a 3D uniform and isothermal turbulent medium. Some of data cubes have been used in \citet{2018arXiv180909806H} and their information based on the representation of Mach numbers and plasma $\beta$ parameter, is listed in Table \ref{tab:Sim}, with numerical resolution of $480^3$. As shown in this table, a wide parameter range for $M_{\rm A}$ and $M_{\rm s}$ is considered. In general, larger $M_{\rm s}$ could be attributed to molecular cloud environments and $M_{\rm s}$ are not too high for high Galactic latitude gas as well. However, some particular environments such as the region from jets of Active Galactic Nuclei interacting with the surrounding ISM still have a large $M_{\rm s}$. 

\begin{table*}
 \centering
 \label{tab:Sim}
 \begin{tabular}{c c c cccc}
Run & $M_{\rm A}$ & $M_{\rm s}$ & $\beta$ & $l_{\rm A}$&$\delta B_{\rm rms}/\left<B\right>$ &Description\\ \hline \hline
A1 & 0.18 & 1.94 & 0.017 &  -- & 0.14&Sub-Alfv{\'e}nic and supersonic \\
A2 & 0.35 & 3.86 & 0.017 & --& 0.26&Sub-Alfv{\'e}nic and supersonic \\
A3 & 0.49 & 0.16 & 18.993 & --& 0.40&Sub-Alfv{\'e}nic and subsonic  \\
A4 & 0.52 & 0.05 & 206.561 &--& 0.39 &Sub-Alfv{\'e}nic and subsonic  \\
A5 & 0.59 & 1.92 & 0.190 & --& 0.46&Sub-Alfv{\'e}nic and supersonic \\
A6 & 0.66 & 7.14 & 0.017&--& 0.48&Sub-Alfv{\'e}nic and supersonic   \\
A7 & 1.08 & 0.10 & 219.238 & --&0.40&Tran-Alfv{\'e}nic and subsonic  \\  
A8 & 1.11 & 0.34 & 20.759 & 93.6& 0.80&Super-Alfv{\'e}nic and subsonic  \\ \hline
B1 & 0.50 & 4.55 & 0.023 & --& 0.37&Sub-Alfv{\'e}nic and supersonic  \\ 
B2 & 0.69 & 5.94 & 0.027 & --& 0.41&Sub-Alfv{\'e}nic and supersonic  \\ 
B3 & 1.08 & 6.38 & 0.057 & -- &0.60&Tran-Alfv{\'e}nic and supersonic  \\ 
B4 & 1.67 & 8.09 & 0.086 & 27.5 &0.81 &Super-Alfv{\'e}nic and supersonic  \\ 
B5 & 2.19 & 5.87 & 0.277 & 12.2&1.16&Super-Alfv{\'e}nic and supersonic  \\ 
B6 & 2.54 & 7.31 & 0.241 & 7.8& 1.08&Super-Alfv{\'e}nic and supersonic  \\ \hline \hline
\end{tabular}
 \caption {Simulations used in our current work with numerical resolution $480^3$. $\beta=2M_{\rm A}^2/M_{\rm s}^2$ is a plasma parameter. $\delta B_{\rm rms}$ indicates the root mean square of random magnetic field and $\left<B\right>$ the regular magnetic field. A1--A8 are from ZEUS simulation and B1--B6 from ATHENA simulation. }\label{Table:SIM}
\end{table*}

\subsection{Synchrotron polarization gradients: single frequency measure}

\begin{figure}
\centering
\includegraphics[width=0.55\textwidth]{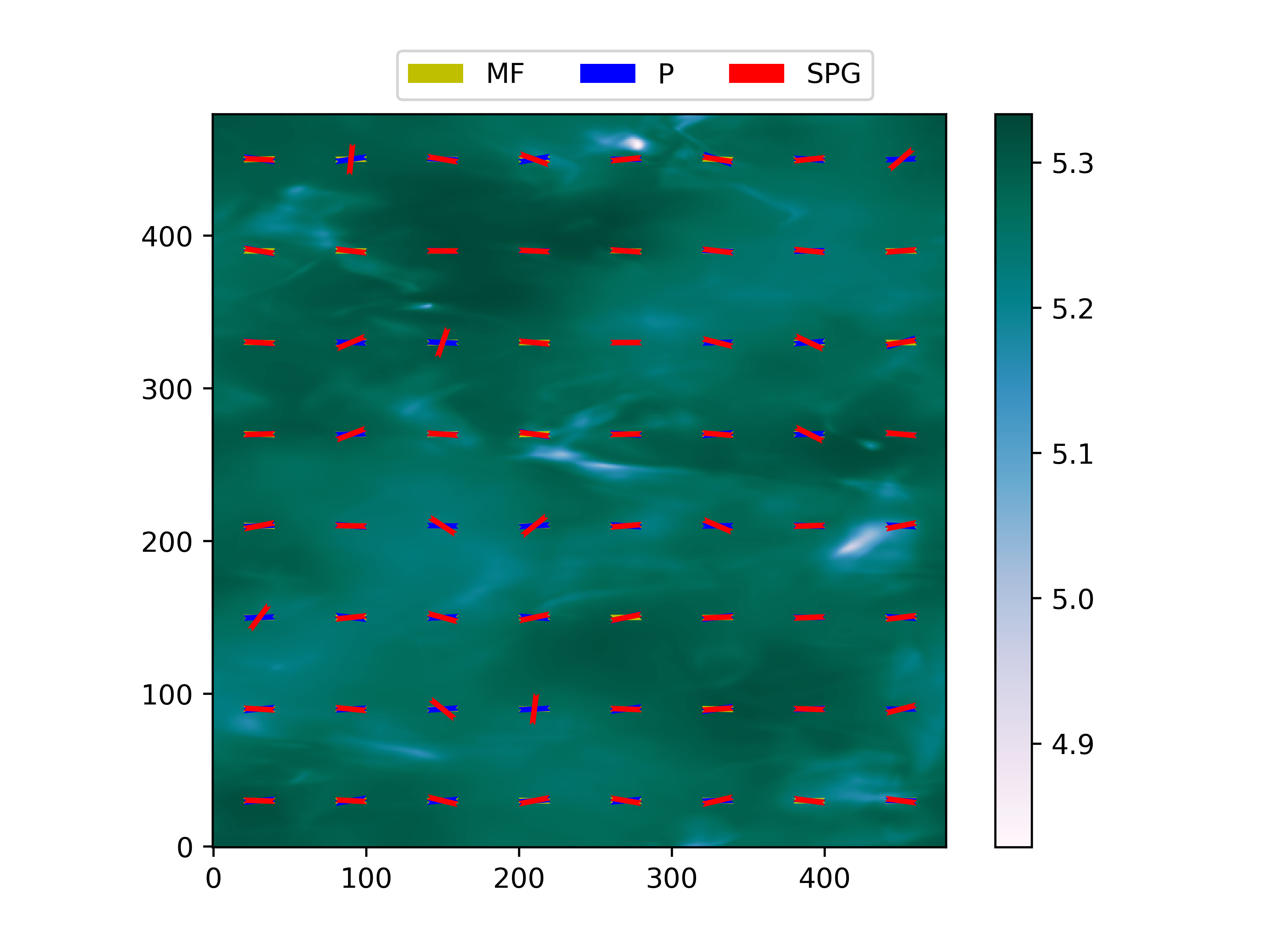}
\caption{\label{fig:map} The rotated $90^\circ$ gradient directions for both SPGs and synchrotron polarization (P), as well as the direction of the projected mean magnetic field (MF) in a synthetic observation obtained by run A1 of Table \ref{tab:Sim}, are overlapped on the background map of synchrotron polarization intensity. The alignment measures between them are $AM$=0.788 (SPGs vs MF), 0.776 (SPGs vs P), and 0.978 (P vs MF).}
\end{figure}

On the basis of LY18, we first explore that the AM of the SPGs versus the direction of mean magnetic field projected in the plane of the sky by using the data A1 of Table \ref{tab:Sim}. Fig. \ref{fig:map} shows that the orientations of SPGs, synchrotron polarization (P) and mean magnetic field (MF) are over-plotted on the map of synchrotron polarization intensity calculated at the frequency of $\nu=1\rm\ GHz$. By using 60-pixel block averaging size, we obtain three AM values as follows:  AM=0.788 (SPGs vs MF), 0.776 (SPGs vs P), and 0.978 (P vs MF). We emphasize that the value of an AM depends on the block size that we set in an integer multiple of numerical resolution. In general, the AM value increases with the block size. However, when comparing the case of the block size of $80$ pixels with that of $60$ pixels, we find that they would increase slowly, which is consistent with those of Fig. 3 in \cite{Lazarian17}. We thus use the block size of $60$ pixels in this study. It can be seen that some regions in Fig. \ref{fig:map} present a bad correspondence between rotated $90^\circ$ gradient directions and magnetic field directions. The reason may be that different (Alfv{\'e}n, slow and fast) MHD turbulence modes are competing with each other and result in an alignment change because the polarization gradients for fast mode has a distinct alignment way with respect to magnetic fields; it will be postponed to a future study.

\begin{figure*}
\centering
\includegraphics[width=0.49\textwidth]{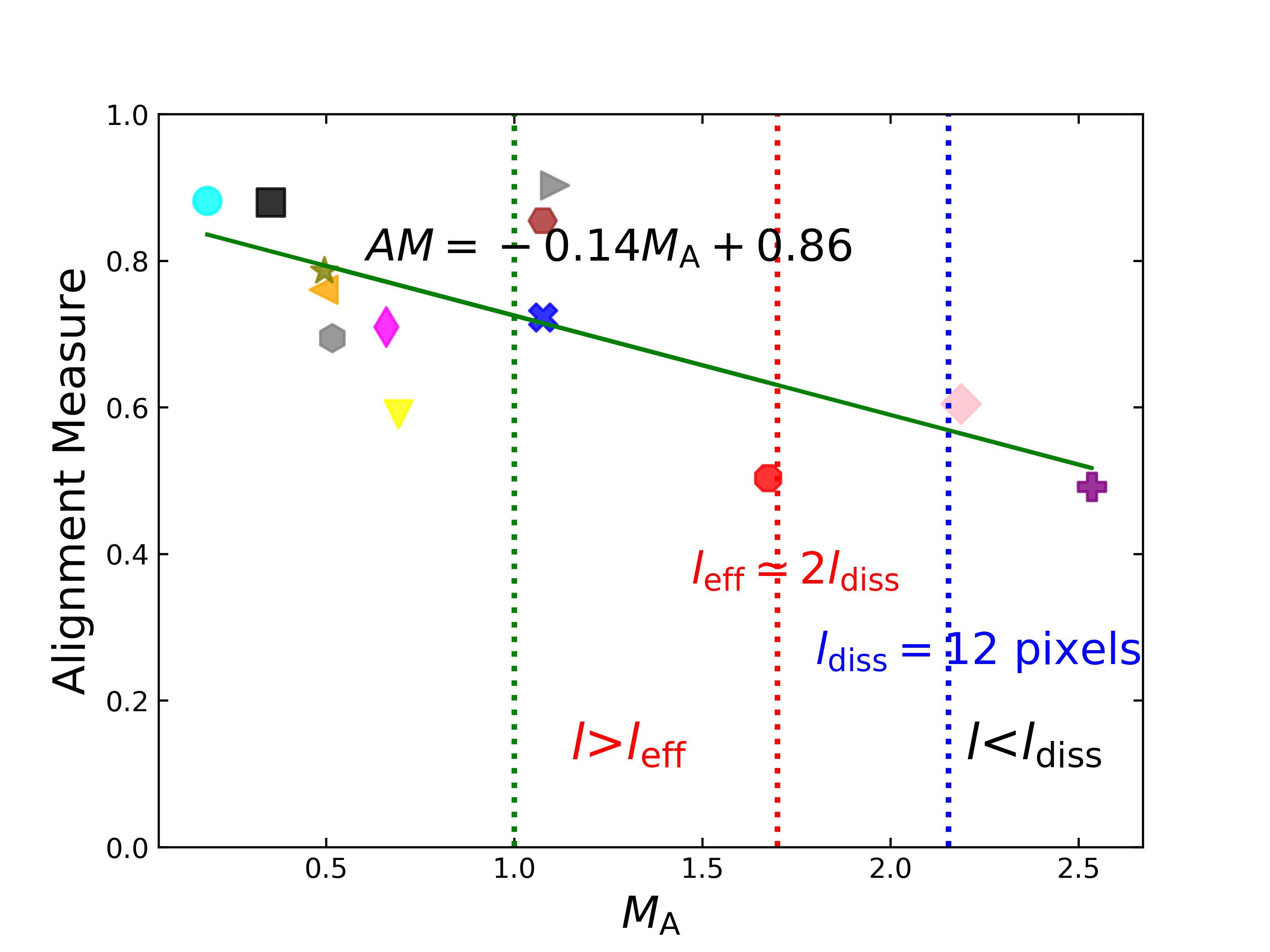}
\includegraphics[width=0.49\textwidth]{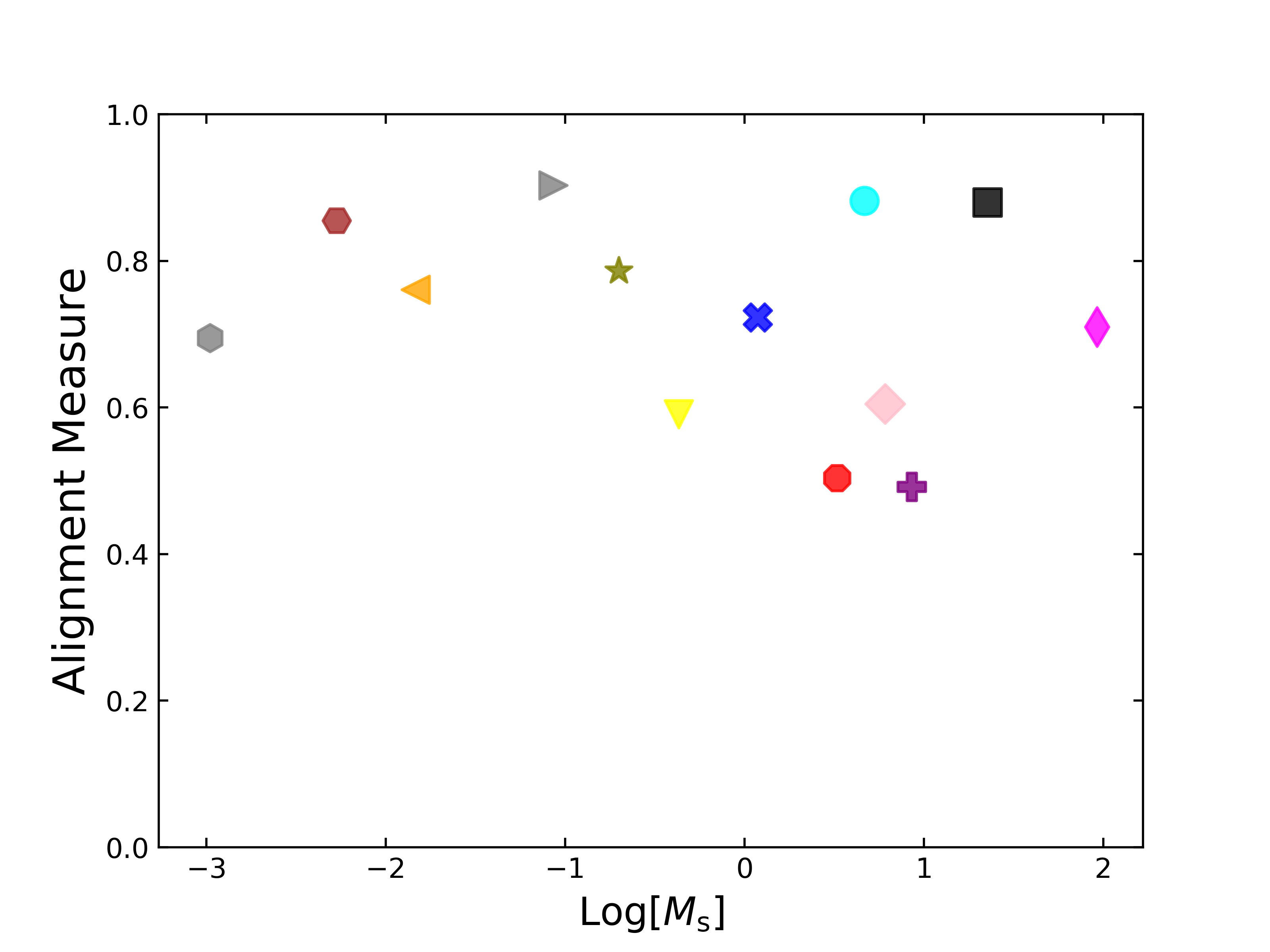}
\caption{\label{fig:AMvsMa} {\it Left panel}: The AM between SPGs and mean magnetic field as a function of $M_{\rm A}$. The solid line is a linear fitting of the AM value. The vertical dotted lines are used to divide super-Alfv{\'e}nic turbulence into different regions: $l> l_{\rm eff}$, $l_{\rm eff} > l > l_{\rm diss}$ and $l < l_{\rm diss}$. {\it Right panel}: The same AM as a function of $M_{\rm s}$. The color symbols have one-to-one correspondence between the left and right panels.
 } 
\end{figure*}

\begin{figure}
\centering
\includegraphics[width=0.5\textwidth]{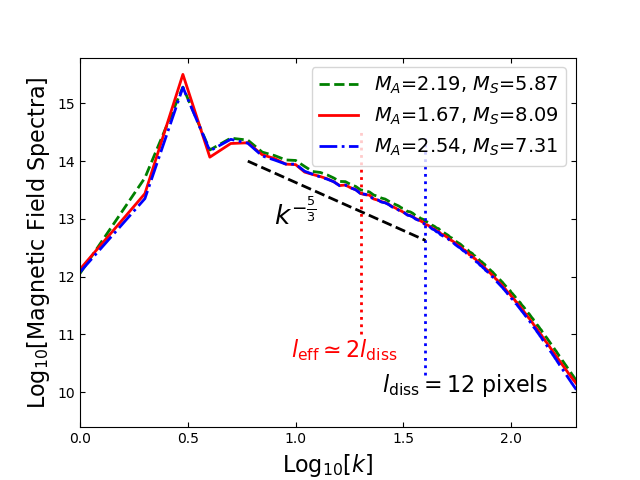}
\caption{\label{fig:AMvsMs} 
 Spectra of magnetic field for super-Alfv{\'e}nic simulations. The vertical dotted lines are used to mark locations of the dissipation scale, $l_{\rm diss}=12\ \rm pixels$, and the approximately transition scale, $l_{\rm eff}\simeq2l_{\rm diss}$, of gradient measurement, in the super-Alfv{\'e}nic region.
 } 
\end{figure}

In order to explore the influence of Mach numbers on the alignment, we here calculate polarization gradients at a very high frequency, i.e., with negligible Faraday depolarization effect. In Fig. \ref{fig:AMvsMa}, we plot the AM between SPGs and magnetic field as a function of $M_{\rm A}$ (left panel) and $M_{\rm s}$ (right panel). Different symbols have a one-to-one correspondence between them. As seen in the right panel, there is a large AM value that provides a good determination of magnetic field directions in different sonic Mach regimes. 

Although we show in the left panel the whole range of $M_{\rm A}$ the magnetic field tracing is expected for $l_{\rm A}>l_{\rm diss}$ from a theoretical point of view (see Section \ref{MHDGMI}). However, when $l_{\rm A}$ is too close to $l_{\rm diss}$ to effectively trace magnetic field, we thus introduce a parameter $l_{\rm eff}$ to quantify an effective region of gradient technique. In our studies, $L_{\rm inj}$ is 128 pixels and $l_{\rm diss}$ approximately 12 pixels. The latter is determined by the cut-off location of the power spectra of magnetic fields plotted in Fig. \ref{fig:AMvsMs} from three super-Alfv{\'e}n simulations, on which we illustrate the dissipation scale, $l_{\rm diss}$, and the effective scale of gradient measures, $l_{\rm eff}\simeq 2l_{\rm diss}$, by the vertical dotted lines. It should be noticed that a specific value of $l_{\rm eff}$ may depend on the numerical resolution. According to equation (\ref{SupLA}), we have $M_{\rm A}\simeq2.2$ at $l_{\rm diss}=12$ pixels. Meanwhile, $M_A$ should be less than 1.7 when $l_{\rm A}>2l_{\rm diss}$. Consequently, we clearly see three regions in Fig. \ref{fig:AMvsMa}, the $l>l_{\rm eff}$ region of which corresponds to the case where gradient techniques work. For high Galactic latitude regions of synchrotron emission we expect $M_{\rm A}\sim 1$ (\citealt{Mao12}) and therefore the AM technique can determine magnetic field directions within the Galactic ISM. 

Formally, a linear fitting of the AM between SPGs and mean magnetic field directions presents a relationship of $AM=-0.14M_{\rm A}+0.86$, which indicates that an AM decreases with increasing $M_{\rm A}$. However, we should not believe to the part for the $M_A>1.7$ and we postpone any actual fitting to the time when we have significantly larger data cubes for which the difference between $L_{\rm inj}$ and $l_{\rm diss}$ is larger enough and therefore the alignment for larger $M_A$ can be tested reliably.

\subsection{Synchrotron polarization gradients: multi-frequency measure}

As for high frequencies polarization method can trace the magnetic field direction well, but for low frequencies the Faraday rotation effect makes the interpretation of the polarization in terms of magnetic field much more complicated. However, the SPGs are expected to robustly trace magnetic field direction in the low frequency range and are not affected by the Faraday rotation. We explore this point by generating synthetic observations corresponding to a wide frequency range

\begin{figure*}
\centering
\includegraphics[width=0.49\textwidth]{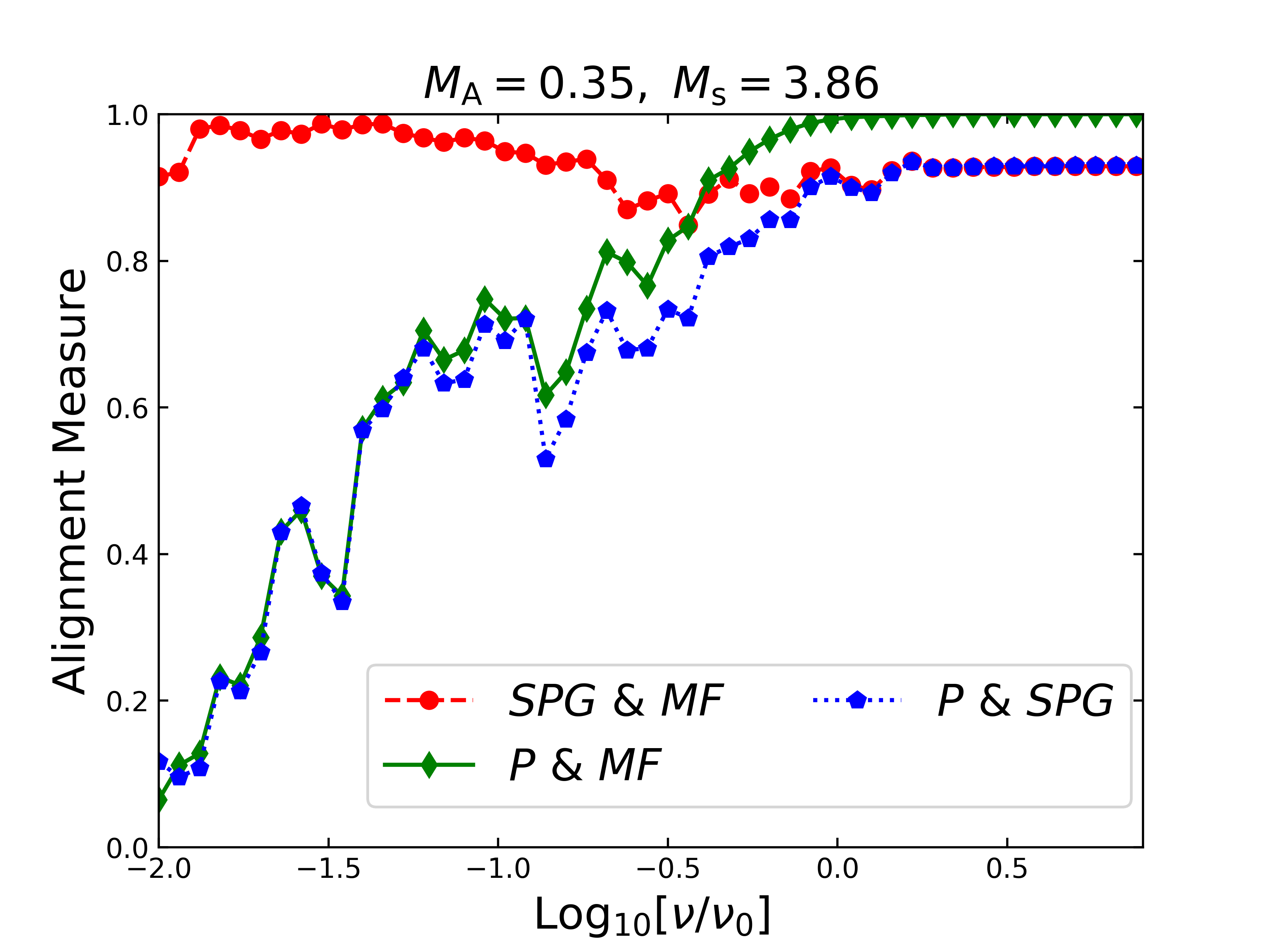}
\includegraphics[width=0.49\textwidth]{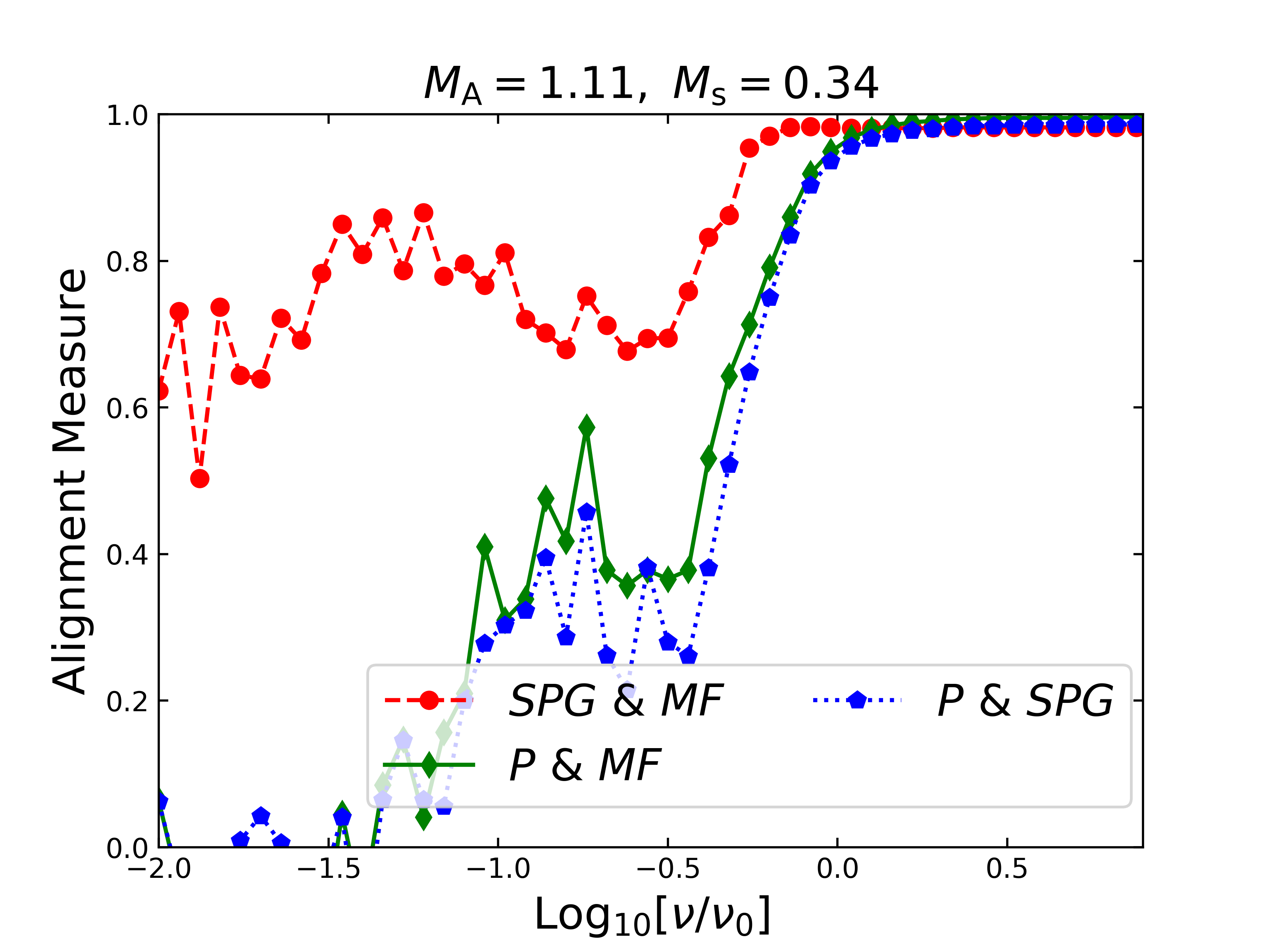}\\
\includegraphics[width=0.49\textwidth]{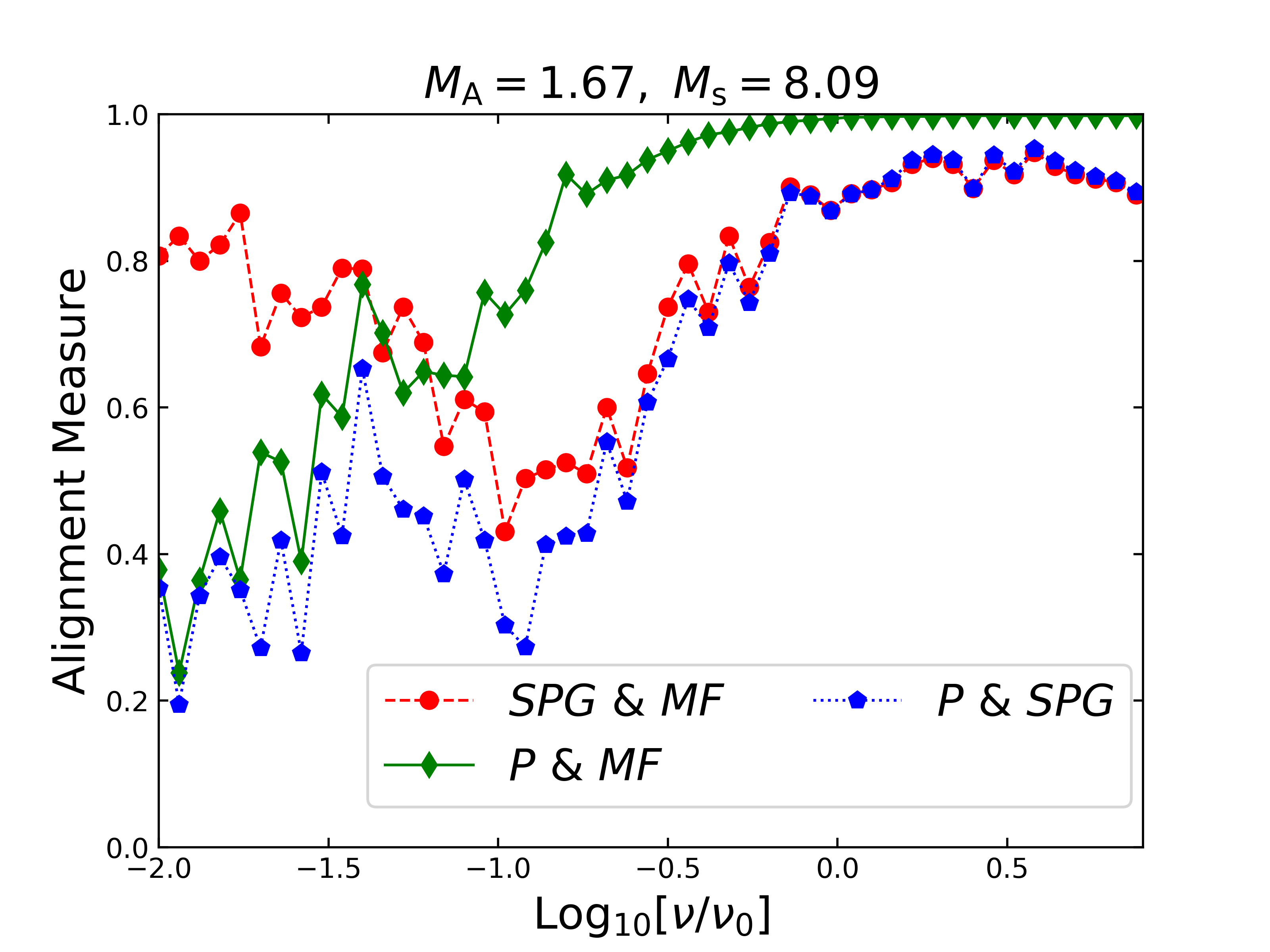}
\includegraphics[width=0.49\textwidth]{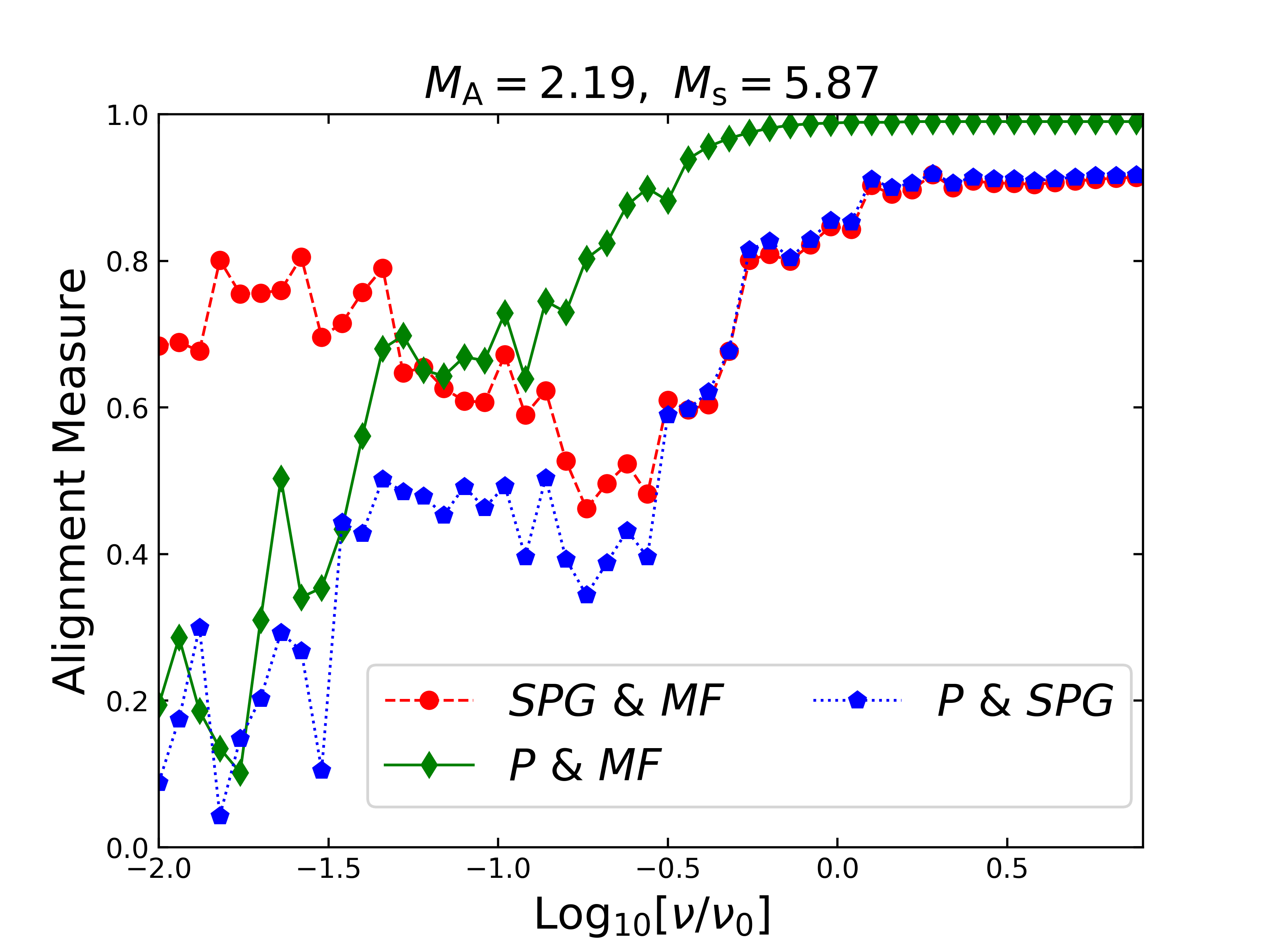}
\caption{\label{fig:AMSPGfreq} The AM as a function of frequency ranging from sub- to super-Alfv{\'e}nic simulations. Small scale, like-noise structures of synthetic synchrotron polarization intensity maps obtained at the frequency of $\leq 0.1~\rm GHz$ are smoothed with $\sigma=2$ pixels Gaussian kernel. Large scale structures of polarization maps are filtered with the approximately wavenumber equal to $1/l_{\rm A}$ of the box size. Sub-Alfv{\'e}n simulation (upper left) does not use any spatial filtering technique. The frequency in the all panels is dimensionless in units of $\nu_0=1 \rm\ GHz$. }
\end{figure*}

In our simulations, the spectral index of relativistic electrons is set as $p=3$ throughout the paper, the change of which does not affect the statistical results of synchrotron emission as is demonstrated in our previous studies (LP12, \citealt{Zhang16,ZhangL18}). Furthermore, the thermal electron density is taken to be $0.01\ \rm cm^{-3}$. The parallel component of the magnetic field is normalized to the mean value of 1 $\mu\rm G$. The AM of SPGs vs magnetic field in a range of multi-frequencies are shown in Fig. \ref{fig:AMSPGfreq}, comparing with the measure of both SPGs vs polarization and polarization vs magnetic field. The left upper panel of Fig. \ref{fig:AMSPGfreq} presents the result for sub-Alfv{\'e}nic and supersonic turbulence with the parameters $M_{\rm A}=0.35$ and $M_{\rm s}=3.68$ (see A2 of Table \ref{tab:Sim}). We can see that the SPGs have a larger AM value than the other two measures at lower frequency bands, where synchrotron polarization emissions suffer from the strong Faraday depolarization, which results in a failure of the traditional polarization measure by $\theta={1\over 2}{\rm actan}(U/Q)$. Super-Alfv{\'e}nic and subsonic simulation is plotted in the right upper panel of Fig. \ref{fig:AMSPGfreq} using $M_{\rm A}=1.11$ and $M_{\rm s}=0.34$ (see A8 of Table \ref{tab:Sim}). At low frequency band ($\nu\lesssim 0.1\rm\ GHz$), we use a full width half maximum of Gaussian kernel of $\sigma=2$ pixels to smooth small scale noise-like structures. We filter large scale spatial structures for the approximately wavenumber equal to $1/l_{\rm A}$ of the box size, following the theoretical guidance of equation (\ref{SupLA}).  

Super-Alfv{\'e}nic and supersonic simulations are shown in the lower panels of Fig. \ref{fig:AMSPGfreq}, in which 3D data cubes are generated using ATHENA code. Similar to the case of the upper right panel of this figure, we use also the spatial filtering technique to filter small scale ( $\sigma=2$ pixels at low frequencies) and large scale structures for excluding the influence of hydrodynamic-like structures (see equation \ref{SupLA}). We find that the SPGs can better measure the direction of the {\it global} projected mean magnetic field at lower frequency and strong Faraday depolarization regions. It can be seen that the curve of the SPGs measure presents a trough-like shape in the intermidiate frequency range, which seems due to an influence of weak random magnetic fields parallel to the line of sight that results in a random-field-dominated Faraday rotation fluctuation (see LP16 for more details). 

\subsection{Synchrotron polarization gradients: influences of mean magnetic field directions }
\subsubsection{Rotation of the cube}\label{RC}
Our earlier studies similar to the studies in LY18 assumed that the mean magnetic field is perpendicular to the line of sight. Such settings are common for the interstellar magnetic field studies, but it is good to study a general case when the magnetic field and the line of sight are at an arbitrary angle. 

\begin{figure*}
\centering
\includegraphics[width=0.95\textwidth]{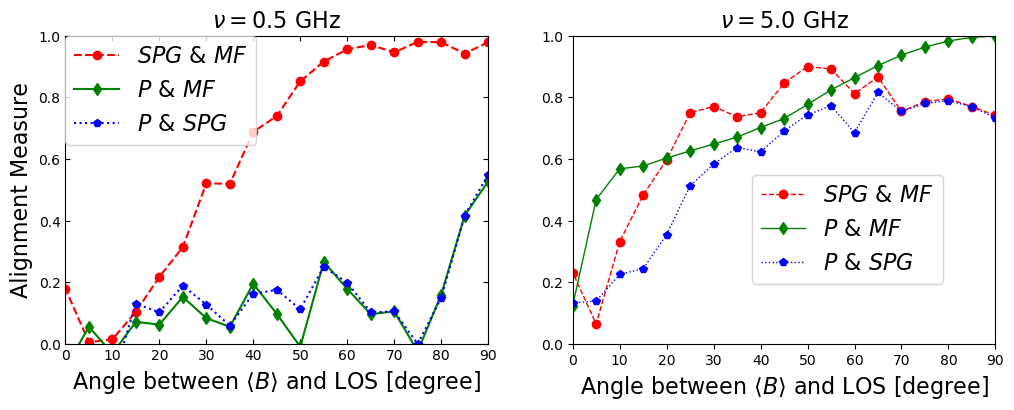}
\caption{\label{fig:rotationY} The AM determined at the frequencies $0.5$ (left panel) and $5\ \rm GHz$ (right panel), as a function of the angle between the line of sight (LOS) and mean magnetic field. These two panels are plotted in terms of  the simulation of A1 in Table \ref{Table:SIM}.}
\end{figure*}

To explore the influence of the direction of mean magnetic fields on the SPGs measure, we rotate the A1 box of Table \ref{tab:Sim} along the $y$-axis located in the plane of the sky, which results in mean magnetic fields to deviate from the plane of the sky. We can see that the AM of the SPGs in Fig. \ref{fig:rotationY} increase with increasing angle between the line of sight and the mean magnetic field direction. The left panel shows the measure calculated at a low frequency of $\nu=0.5\rm\ GHz$, where there exists a strong Faraday depolarization. We find that the polarization cannot trace the magnetic field direction but the SPGs can surely provide a tracing for magnetic field distribution. At a high frequency, $\nu=5\rm\ GHz$ (right panel), it shows that all the AMs increase with angle between the line of sight and the mean magnetic field due to a weakened Faraday rotation. It is very evident that the SPGs can better trace the projected mean magnetic fields in the plane of the sky. 

\begin{figure*}
\centering
\includegraphics[width=0.95\textwidth]{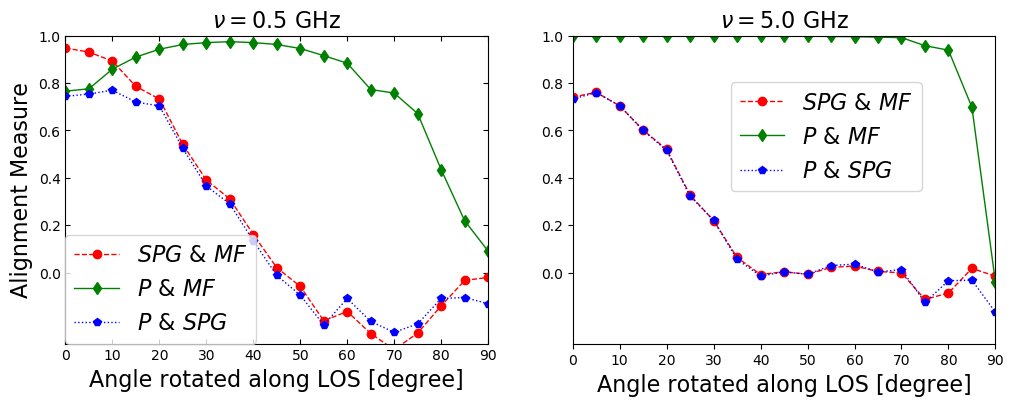}
\caption{\label{fig:rotationX} The AM determined at the frequencies $0.5$ (left panel) and $5\ \rm GHz$ (right panel), as a function of rotated angle of back half cube along the line of sight. The two panels are based on the simulation of A1 in Table \ref{Table:SIM}. }
\end{figure*}

\subsubsection{Magnetic field changing direction along the line of sight}
Now, we divide the same box used in Section \ref{RC} into two equivalent parts and rotate one of them along the line of sight. The resulting plots are presented in Fig. \ref{fig:rotationX}, which is calculated at the frequencies of $\nu=0.5$ (left panel) and $\nu=5\rm\ GHz$ (right panel), respectively. It can be found that the AMs decrease with increasing the rotated angles. When rotating up to the maximum $90^\circ$, that is, the mean magnetic field directions within two parts of the box is perpendicular each other, all the AMs cannot trace the direction of mean magnetic fields. As will be described in Section \ref{SPDGTM}, but one can still sample the direction of the magnetic field within the closer region by changing the frequency at the lower frequency region in terms of the SPDGs technique. 

\subsection{Synchrotron polarization derivative gradients: tomographic measure}\label{SPDGTM}

Our studies so far dealt with the SPGs. The other measure, namely, the SPDGs technique was also introduced in LY18. It is more sensitive to Faraday rotation and can be used to study the magnetic field in regions, e.g. with higher density of thermal electrons. In this sense, this measure is complementary to the SPGs.

\begin{figure}
\centering
\includegraphics[width=0.5\textwidth]{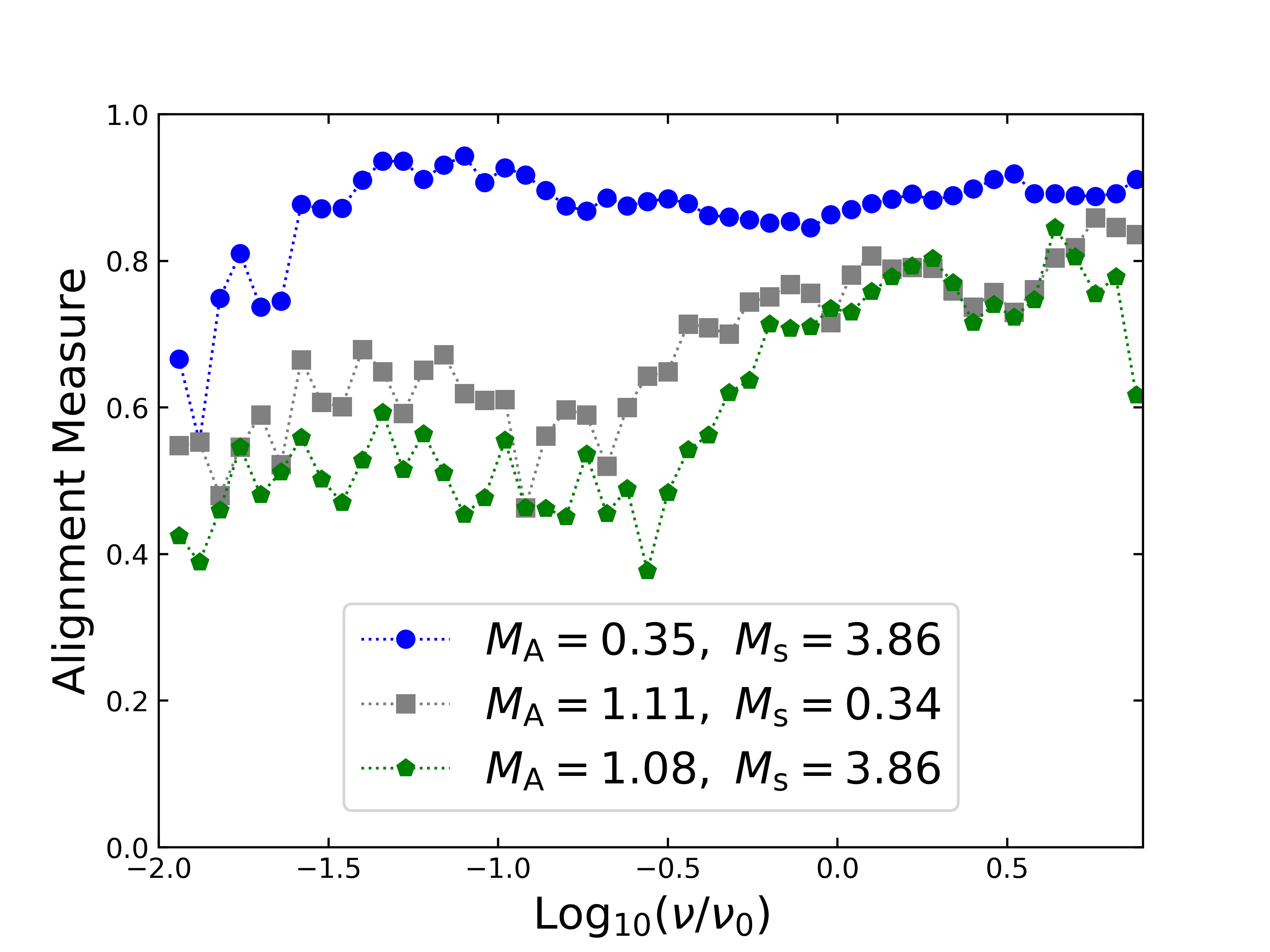}
\caption{\label{fig:SPDGvsMF} The AM between the SPDGs and local mean magnetic fields as a function of frequency. Super-Alfv{\'e}n simulations use a filtering technique that is the same as in Fig. \ref{fig:AMSPGfreq}. }
\end{figure}

In Fig. \ref{fig:SPDGvsMF}, we present the results for the AM of the SPDGs, which traces the direction of the {\it local} mean magnetic fields ranging from sub- to super-Alfv{\'e}nic turbulence regions. According to equation (\ref{eq:el}), we know that in the strong Faraday rotation regime, the SPGs sample magnetic fields up to a depth of $L_{\rm eff}$ (see Section \ref{SynTheor}), the tracing directions of which are not distorted by the Faraday depolarization unlike the polarization method. Because $L_{\rm eff}$ depends on the squared wavelength $\lambda^2$, one can sample the magnetic field information at different distances by changing the wavelength. If choosing two frequency points $\nu_{\rm 1}$ and $\nu_{\rm 2}$ with an interval $\Delta \nu=\nu_{\rm 2}-\nu_{\rm 1}$, one has
$\Delta \lambda^2=c^2\Delta \nu / \nu^3$. We can thus obtain the synchrotron polarization intensity derivative with regard to the squared wavelength $\lambda^2$ following equation (\ref{eq:SPDG}).

As shown in Fig. \ref{fig:SPDGvsMF}, the SPDGs can extract the local mean magnetic fields within a Faraday depth interval $\Delta L=\Delta\nu^2/(c^2\phi)$, ranging from sub- to super-Alfv{\'e}nic turbulence regions. The main purpose of our current simulation is to confirm the feasibility of our SPDGs to trace magnetic field and we do not focus on the technical improvement of measure methods. We expect that the level of the AM can be increased further if additional techniques, such as angle constraint and moving window, are employed (see \citealt{LY18b}).

\begin{figure}
\centering
\includegraphics[width=0.5\textwidth]{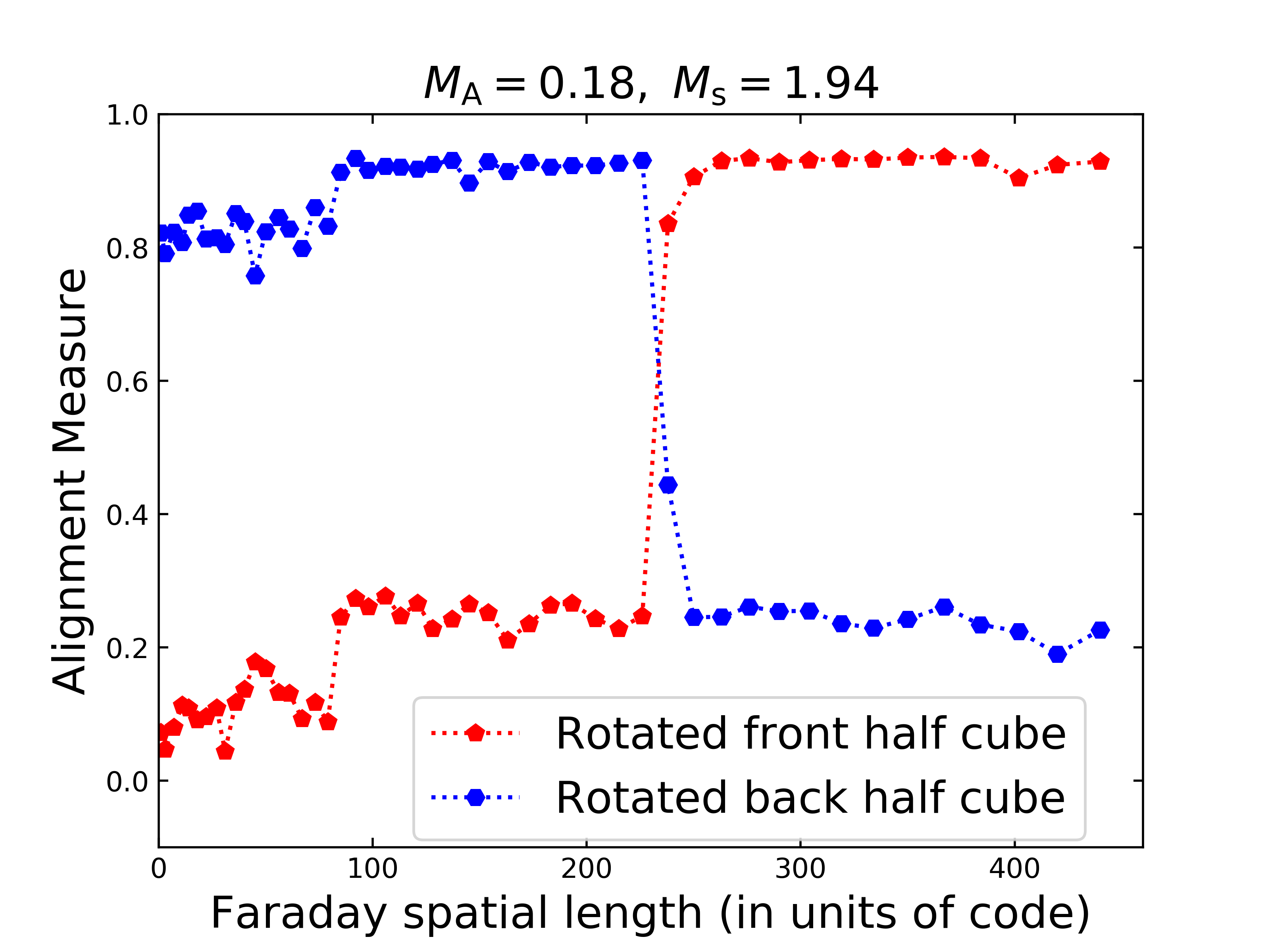}
\caption{\label{fig:rotationXFB} The AM as a function of Faraday rotation spatial extent. The results of AM for the front and back half cubes are based on the simulation A1 in Table \ref{Table:SIM}. The radiation frequency is set in the range from 0.1 to $0.25\ \rm GHz$. }
\end{figure}

To further figure out the tomographic studies of magnetic fields by the SPDGs, we divide the box A1 listed in Table \ref{tab:Sim} into two equivalent parts along the line of sight. Then one of them is rotated respectively $30^\circ$ along the line of sight but the other part remains the same. The results of the AM as a function of Faraday rotation real depth in units of code are plotted in Fig. \ref{fig:rotationXFB}, which corresponds to the frequency in the range of $\nu=0.1 - 2.5\ \rm GHz$. For the case of the rotated front half box, we obtain a larger AM value in the back half part of the box compared to the front half box. We see also a similar behavior when rotating the back half cube. 

When we rotate an angle $\theta'$ for one part of the box, the perpendicular component, $B_x$, of mean magnetic field along the $x$-aixs would change into $B_x{\rm cos}\theta'$. As a result of rotation, the projected magnetic field components become weaker, which results in
decreasing in the AM values. Therefore, this exploration has demonstrated that the SPDGs can successfully extract the direction of local magnetic fields by changing the observational frequency. We can sample the magnetic field direction in the closer observer region at lower frequencies, which would provide us an important contribution for reconstructing Galactic 3D magnetic fields. 

\section{Discussions}
One of advantages of using numerical simulation to study MHD turbulence is that we can know the distribution of the turbulent magnetic field in advance. The directions of the SPGs and SPDGs can be directly compared to magnetic field directions, i.e., the AM between the rotated $90^\circ$ synchrotron gradients and magnetic field directions. The SPGs can be used to trace the global mean magnetic field distributions in the plane of the sky, and the SPDGs can measure the direction of the local mean field within a slice of Faraday tomography. The former is mostly sensitive to the synchrotron polarization intensity fluctuation, while the latter is mainly depended on the polarization fluctuation arising from the Faraday rotation density, which provides a very promising method to reconstruct 3D magnetic fields in the Milky Way and external galaxies.   

 As shown in Fig. \ref{fig:AMvsMa},  the ability of the traditional polarization measure tracing magnetic field would be reduced with increasing Alfv{\'e}nic Mach number, which thus is challenging for recovering MHD turbulent properties in the super-Alfv{\'e}n region. Our investigation demonstrated that the SPGs and SPDGs can be applied to reveal the turbulence properties from sub- to super-Alfv{\'e}n regions. However, we need to consider more factors that could impede the alignment determination, such as small scale, noise-like structures, and larger scale structures than $l_{\rm A}$ (see equation \ref{SupLA}). If $M_{\rm A}$ is sufficiently large the scale $l_{\rm A}$ will become small. When $l_{\rm A}$ is comparable to the small dissipation scale $l_{\rm dis}$, the inertial range will be entirely eliminated. Therefore, the studies of super-Alfv{\'e}nic turbulence is subject to a numerical resolution; the high numerical resolution scenario will be explored in the future.  

Our studies are focused on the multi-frequency and strong Faraday depolarization regimes, which is to prepare for the application of techniques to the Low Frequency Array for Radio astronomy (LOFAR) and the Square Kilometer Array (SKA) data. At very high frequency bands, synchrotron polarization emissions suffer from a weak Faraday rotation, where the SPGs would determine the properties of synchrotron emitting regions. This is similar to what the SIGs studied in \cite{Lazarian17} can do. They have reported that SIGs is a reliable technique for tracing magnetic field, which is not subject to a Faraday depolarization. In this paper, we have combined both SPGs and SPDGs with measurements of polarization to obtain a synergy of the different measurements. 

The synchrotron polarization gradient techniques we studied in this paper can be viewed as a part of the gradient technique that is applied to obtain insight into the nature of MHD turbulence (eg., GS95, LV99, \citealt{Cho2002PRL}). Based on the fact that velocity eddies are elongated along local magnetic field directions (\citealt{Gold95}), the gradients of velocities were developed by measuring the velocity centroid gradients (VCG: \citealt{GL17,YL17,Hu2018}). Recently, this measure is further developed to include the velocity channel gradients and the reduced velocity centroid gradients (\citealt{LY18b}). Meanwhile, the VCGs have been applied to measure the magnetization of the ISM by relating the dispersions of velocity gradients to the Alfv{\'e}nic Mach number $M_{\rm A}$ (\citealt{LazarianYH18}). It was claimed in LY18 that the same procedures are valid for obtaining $M_A$ with synchrotron gradients. We are going to demonstrate this point in another paper. 

Very recently, a comparative study between traditional Faraday rotation synthesis technique (see \citealt{Burn66}) and synchrotron gradients showed that in the situations when the traditional technique fails for the case of insufficient frequency coverage, the synchrotron gradients can still trace the 3D distribution of magnetic field (\citealt{Ho2019}). In addition, we would like to mention that the gradient technique is complementary to correlation function anisotropies (\citealt{LP12}) for studying magnetic fields. \cite{Lazarian17} claimed that for synchrotron intensities the latter can correctly trace mean magnetic field, but fails to trace the detailed magnetic field structure. A more detailed study using the velocity gradients and velocity anisotropies in \cite{Yuen18} confirmed this. 

The current work neglects the influence of polarization synchrotron self-absorption on the AM. In general, Faraday rotation would take effect in the relatively low frequency bands, where the synchrotron self-absorption effect should be important for some astrophysical environments. We expect that the effects of self-absorption can provide yet another method of testing the 3D structure of magnetic field. Indeed, only the regions closer to the observer and less affected by self-absorption are expected to be probed under strong self-absorption. This effect is similar to the effect of the dust self-absorption in spectral line statistics that was explored analytically in \cite{Kandel17} and numerically in \cite{2018arXiv180909806H}.

We would also like to emphasize that our gradient measurement is a developing new technique tracing the direction and strength of the magnetic field in the diffuse media. The accurate measurement level of the synchrotron polarization gradient would be further enhanced by optimizing fitting ways and improving sub-block averaging method, for instance, combining the {\it k-means} cluster analysis algorithm (e.g., \citealt{Sanchhez10} ) to divide sub-block of an image.

\section{Summary}
This work has promoted the gradient technique research of the synchrotron polarized radiation carried out in LY18 to super-Alfv{\'e}nic turbulence. We focused on the multi-frequency AM of the SPGs and tomographic analysis of the SPDGs and explore how to recover the underlying magnetic field directions. The main results that we obtained are summarized as follows:

\begin{enumerate}[wide, labelwidth=!, labelindent=0pt]
\item We have found that the AM of the SPGs decreases with increasing Alfv{\'e}nic Mach numbers. Our studies are limited to $M_{\rm A}<1.7$ due to the limited numerical resolution of our simulations. This, however, is not so restrictive as $M_{\rm A}\sim1$ is expected at the high latitude regions of the Milky Way. 

\item At the low frequency and strong Faraday rotation region, the SPGs have a significant advantage over the traditional polarization method in tracing projected mean magnetic fields. Therefore, the SPGs can be employed to correct the magnetic field direction obtained by polarization one.  

\item The SPGs and SPDGs have been successfully applied to trace the global and local mean magnetic fields, respectively. The latter has a unique advantage for reconstructing 3D magnetic fields.

\item Our simulations have demonstrated that synchrotron gradient techniques can recover the underlying magnetic fields, which has taken the first step for applying these techniques to a large number of data set, e.g., from the LOFAR and SKA.

\end{enumerate}

\section*{ACKNOWLEDGMENTS}
We thank an anonymous referee for useful comments that improved the manuscript. We thank Siyao Xu for useful discussions. J.F.Z. thanks the supports from the National Natural Science Foundation of China (grant No. 11703020) and the Hunan Provincial Natural Science Foundation (grant No. 2018JJ3484). A.L. acknowledges the support of NSF grant DMS 1622353 and AST 1816234.




\begin{thebibliography}{99}
\providecommand\natexlab[1]{#1}
\providecommand\JournalTitle[1]{#1}

\bibitem[Beck et al.(2012)]{Beck12} Beck R., Frick P., Stepanov R., \& Sokoloff D.,\ 2012, 
\href{http://dx.doi.org/10.1051/0004-6361/201219094}{\JournalTitle{A\&A}, 543, A113} 

\bibitem[Beck \& Wielebinski (2013)]{Beck13} Beck R., \& Wielebinski R.,\ 2013, Planets, Stars and Stellar Systems.~Volume 5: Galactic Structure and Stellar Populations, 
\href{http://dx.doi.org/10.1007/978-94-007-5612-0_13} {5, 641}

\bibitem[Brentjens \& de Bruyn (2005)]{Brentjens05}Brentjens M., \& de Bruyn A., 2005
\href{http://dx.doi.org/10.1051/0004-6361:20052990}{\JournalTitle{A\&A}, 441, 1217}

\bibitem[Burkhart, Lazarian \& Gaensler (2012)]{Burkhart12} Burkhart B., Lazarian A., \& Gaensler B.~M.,\ 2012, 
\href{http://dx.doi.org/10.1088/0004-637X/749/2/145}{\JournalTitle{ApJ}, 749, 145}

\bibitem[Burn (1966)]{Burn66}Burn B. J., 1966, 
\href{http://dx.doi.org/10.1093/mnras/133.1.67}{\JournalTitle{MNRAS}, 133, 67}

\bibitem[Cho \& Vishniac (2000)]{ChoV00} Cho J., \& Vishniac E.~T.,\ 2000, 
\href{http://adsabs.harvard.edu/abs/2000ApJ...539..273C} {\JournalTitle{ApJ},  539, 273} 

\bibitem[Cho, Lazarian \& Vishniac (2003)]{Cho2003LNP} Cho J., Lazarian A., \& Vishniac E.~T.,\ 2003, Turbulence and Magnetic Fields in Astrophysics, 
\href{http://adsabs.harvard.edu/abs/2003LNP...614...56C}{\JournalTitle{LNP}, 614, 56}

\bibitem[Cho, Lazarian, \& Vishniac (2002)]{ChoLV02} Cho J., Lazarian A., \& Vishniac E.~T.,\ 2002,
\href{http://adsabs.harvard.edu/abs/2002ApJ...564..291C} {\JournalTitle{ApJ}, 564, 291}

\bibitem[Cho \& Lazarian (2003)]{ChoL03} Cho J., \& Lazarian A.,\ 2003, 
\href{http://adsabs.harvard.edu/abs/2003MNRAS.345..325C} {\JournalTitle{MNRAS}, 345, 325} 

\bibitem[{Cho \& Lazarian (2002)}]{Cho2002PRL}
Cho J., \& Lazarian A., 2002,
  \href{http://dx.doi.org/10.1103/PhysRevLett.88.245001}{\JournalTitle{PhRvL}, 88, 245001}

\bibitem[Dickey et al. (2018)]{Dickey18} Dickey J.~M., Landecker T.~L., Thomson A.~J.~M., et al.,\ 2018, arXiv:1812.05399 

\bibitem[Elmegreen \& Scalo (2004)]{Elmegreen04} Elmegreen B.~G., \& Scalo J.,\ 2004, 
\href{http://dx.doi.org/10.1146/annurev.astro.41.011802.094859}{\JournalTitle{ARA\&A}, 42, 211} 

\bibitem[Gaensler et al. (2011)]{Gaensler11} Gaensler B.~M., Haverkorn M., Burkhart B., et al.,\ 2011,
\href{http://dx.doi.org/10.1038/nature10446} {\JournalTitle{Nature}, 478, 214 }

\bibitem[Galtier et al. (2000)]{Galtier00} Galtier S., Nazarenko S.~V., Newell A.~C., \& Pouquet A.,\ 2000,
\href{http://adsabs.harvard.edu/abs/2000JPlPh..63..447G} {\JournalTitle{JPlPh} 63, 447} 


\bibitem[{Goldreich \& Sridhar (1995)}]{Gold95}
Goldreich P.~ \& Sridhar S., 1995,
\href{http://dx.doi.org/10.1086/174600}{\JournalTitle{ApJ}, 438, 763}
  
\bibitem[Gonz{\'a}lez-Casanova \& Lazarian (2017)]{GL17} Gonz{\'a}lez-Casanova D.~F., \& Lazarian A.,\ 2017,
\href{http://adsabs.harvard.edu/abs/2017ApJ...835...41G}{\JournalTitle{ApJ}, 835, 41}

\bibitem[Guo et al. (2017)]{Guo17} Guo X., Mao J., \& Wang J.,\ 2017,
\href{http://dx.doi.org/10.3847/1538-4357/aa7385}{\JournalTitle{ApJ}, 843, 23} 

\bibitem[Fletcher et al. (2011)]{Fletcher11} Fletcher A., Beck R., Shukurov A., Berkhuijsen E.~M., \& Horellou C.,\ 2011, 
\href{http://dx.doi.org/10.1111/j.1365-2966.2010.18065.x} {\JournalTitle{MNRAS}, 412, 2396}

\bibitem[Frick et al. (2011)]{Frick11} Frick P., Sokoloff D., Stepanov R., \& Beck R.,\ 2011,  
\href{http://dx.doi.org/10.1111/j.1365-2966.2011.18571.x}{\JournalTitle{MNRAS}, 414, 2540}

\bibitem[Haverkorn (2015)]{Haverkorn15} Haverkorn M.,\ 2015, Magnetic Fields in Diffuse Media, 407, 483 

\bibitem[Hayes et al. (2006)]{Hayes06} Hayes J.~C., Norman M.~L., Fiedler R.~A., et al.,\ 2006,
\href{http://dx.doi.org/10.1086/504594}{\JournalTitle{ApJS}, 165, 188 }

\bibitem[Herron et al. (2016)]{Herron16} Herron C.~A., Burkhart B., Lazarian A., Gaensler B.~M., \& McClure-Griffiths N.~M.,\ 2016, 
\href{http://dx.doi.org/10.3847/0004-637X/822/1/13}{\JournalTitle{ApJ}, 822, 13}

\bibitem[Herron et al. (2017)]{Herron17} Herron C.~A., Federrath C., Gaensler B.~M., et al.,\ 2017, 
\href{http://dx.doi.org/10.1093/mnras/stw3319}{\JournalTitle{MNRAS}, 466, 2272} 

\bibitem[Herron et al. (2018a)]{Herron18a} Herron C.~A., Burkhart B., Gaensler B.~M., et al.,\ 2018a,
\href{http://dx.doi.org/10.3847/1538-4357/aaafd0} {\JournalTitle{ApJ}, 855, 29} 

\bibitem[Herron et al. (2018b)]{Herron18b} Herron C.~A., Gaensler B.~M., Lewis G.~F., \& McClure-Griffiths N.~M.,\ 2018b, 
 \href{http://dx.doi.org/10.3847/1538-4357/aaa002}{\JournalTitle{ApJ}, 853, 9 }

\bibitem[Ho et al.(2019)]{Ho2019} Ho, K.~W., Yuen, K.~H., Leung, P.~K., \& Lazarian, A.,\ 2019, arXiv:1901.07731 

\bibitem[Hu et al. (2018)]{Hu2018} Hu Y., Yuen K.~H., \& Lazarian A.,\ 2018, MNRAS, 480, 1333 

\bibitem[Hsieh et al. (2018)]{2018arXiv180909806H} Hsieh, C.-H., Hu, Y., Lai, S.-P., et al.,\ 2019, ApJ, 873, 16 

\bibitem[Kowal \& Lazarian(2010)]{Kowal10}Kowal G., \& Lazarian A.,\ 2010,
 \href{http://adsabs.harvard.edu/abs/2010ApJ...720..742K}{\JournalTitle{ApJ}, 720, 742} 

\bibitem[Iacobelli et al. (2014)]{Iacobelli14} Iacobelli M., Burkhart B., Haverkorn M., et al.,\ 2014, 
\href{http://dx.doi.org/10.1051/0004-6361/201322982}{\JournalTitle{A\&A}, 566, A5} 

\bibitem[Jeli{\'c} et al. (2015)]{Jelic15} Jeli{\'c} V., de Bruyn A.~G., Pandey V.~N., et al.,\ 2015, 
\href{http://dx.doi.org/10.1051/0004-6361/201526638}{\JournalTitle{A\&A}, 583, A137} 

\bibitem[Kandel, Lazarian \& Pogosyan (2017)]{Kandel17} Kandel D., Lazarian A., \& Pogosyan D.\ 2017,
 \href{http://dx.doi.org/10.1051/10.1093/mnras/stx1358} {\JournalTitle{MNRAS}, 470, 3103} 

\bibitem[Kowal et al. (2009)]{Kowal09} Kowal G., Lazarian A., Vishniac E.~T., \& Otmianowska-Mazur K.,\ 2009,  
\href{http://dx.doi.org/10.1088/0004-637X/700/1/63}{\JournalTitle{ApJ}, 700, 63}

\bibitem[Lazarian et al. (2012)]{Lazarian12} Lazarian A., Vlahos L., Kowal G., et al.,\ 2012, 
\href{http://dx.doi.org/10.1007/s11214-012-9936-7}{\JournalTitle{ Space Sci. Rev.}, 173, 557} 

\bibitem[Lazarian et al. (2015)]{Lazarian15} Lazarian A., Eyink G., Vishniac E., \& Kowal G.,\ 2015, 
\href{http://adsabs.harvard.edu/abs/2015RSPTA.37340144L}{\JournalTitle{RSPTA}, 373, 20140144L }

\bibitem[Lazarian (2006)]{Lazarian06} Lazarian A.\ 2006,  
\href{http://dx.doi.org/10.1086/505796}{\JournalTitle{ApJL}, 645, L25}

\bibitem[Lazarian \& Vishniac (1999)]{LazarianV99}
Lazarian A., \& Vishniac E.~T., 1999,
  \href{http://dx.doi.org/10.1086/307233}{\JournalTitle{ApJ}, 517, 700}

\bibitem[{{Lazarian} \& {Pogosyan} (2012)}]{LP12} {Lazarian} A., \& {Pogosyan} D., 2012,
  \href{http://dx.doi.org/10.1088/0004-637X/747/1/5}{\JournalTitle{ApJ}, 747, 5} (LP12)

\bibitem[{{Lazarian} \& {Pogosyan} (2016)}]{LP16}  {Lazarian} A., \& {Pogosyan} D.,\ 2016,
  \href{http://dx.doi.org/10.3847/0004-637X/818/2/178}{\JournalTitle{ApJ}, 818, 178} (LP16)

\bibitem[{{Lazarian} (2007)}]{Lazarian07}
{Lazarian} A., 2007,
  \href{http://dx.doi.org/10.1016/j.jqsrt.2007.01.038}{\JournalTitle{ RSPTA},
  106, 225}
  
 \bibitem[Lazarian \& Yuen (2018a)]{LY18} Lazarian A., \& Yuen K.~H.,\ 2018a,
\href{http://dx.doi.org/10.3847/1538-4357/aad3ca}{\JournalTitle{ApJ}, 865, 59}
  
 \bibitem[Lazarian \& Yuen (2018b)]{LY18b} Lazarian A., \& Yuen K.~H.,\ 2018b, 
 \href{http://dx.doi.org/10.3847/1538-4357/aaa241}{ \JournalTitle{ApJ}, 853, 96} 
 
 \bibitem[Lazarian et al. (2018)]{LazarianYH18} Lazarian A., Yuen K.~H., Ho K.~W., et al.,\ 2018, 
 \href{http://dx.doi.org/10.3847/1538-4357/aad7ff}{ \JournalTitle{ApJ}, 865, 46} 

\bibitem[Lazarian (2006)]{Lazarian06} Lazarian A.,\ 2006,
  \href{http://dx.doi.org/10.1086/505796}{\JournalTitle{ApJL}, 645, L25 }

\bibitem[Lazarian et al. (2017)]{Lazarian17} Lazarian A., Yuen K.~H., Lee H., \& Cho J.,\ 2017, 
\href{http://dx.doi.org/10.3847/1538-4357/aa74c6}{\JournalTitle{ApJL}, 842, 30 }

\bibitem[Lee et al. (2016)]{Lee16} Lee H., Lazarian A., \& Cho J.,\ 2016, 
\href{http://dx.doi.org/10.3847/0004-637X/831/1/77}{\JournalTitle{ApJ}, 831, 77 }

\bibitem[Lithwick \& Goldreich(2001)]{Lithwick01} Lithwick Y., \& Goldreich P.,\ 2001,
\href{http://adsabs.harvard.edu/abs/2001ApJ...562..279L}{\JournalTitle{ApJ}, 562, 279} 

\bibitem[Mao et al.(2012)]{Mao12} Mao S.~A., McClure-Griffiths N.~M., Gaensler B.~M., et al.,\ 2012, 
\href{http://adsabs.harvard.edu/abs/2012ApJ...755...21M}{\JournalTitle{ApJ}, 755, 21}

\bibitem[Maron \& Goldreich (2001)]{Maron01} Maron J., \& Goldreich P.,\ 2001, 
\href{http://adsabs.harvard.edu/abs/2001ApJ...554.1175M} {\JournalTitle{ApJ}, 554, 1175 }

\bibitem[McKee \& Ostriker (2007)]{Mckee07} McKee C.~F., \& Ostriker E.~C.,\ 2007,  
\href{http://dx.doi.org/10.1146/annurev.astro.45.051806.110602}{\JournalTitle{ARA\&A}, 45, 565 }

\bibitem[Narayan \& Medvedev (2001)]{Narayan01} Narayan R., \& Medvedev M.~V.,\ 2001, 
\href{http://dx.doi.org/10.1086/338325}{\JournalTitle{ApJL}, 562, L129} 

\bibitem[Robitaille \& Scaife (2015)]{Robitaille15} Robitaille J.-F., \& Scaife A.~M.~M.,\ 2015, 
\href{http://dx.doi.org/10.1093/mnras/stv920}{\JournalTitle{MNRAS}, 451, 372} 

\bibitem[S{\'a}nchez Almeida et al.(2010)]{Sanchhez10} S{\'a}nchez Almeida J., Aguerri J.~A.~L., Mu{\~n}oz-Tu{\~n}{\'o}n C., \& de Vicente A.,\ 2010, 
\href{http://adsabs.harvard.edu/abs/2010ApJ...714..487S}{\JournalTitle{ApJ}, 714, 487} 

\bibitem[Schlickeiser (2002)]{Schlickeiser02}Schlickeiser R., 2002, Cosmic Ray Astrophysics (Berlin: Springer)

\bibitem[Stone et al. (2008)]{Stone08} Stone J.~M., Gardiner T.~A., Teuben P., Hawley J.~F., \& Simon J.~B.,\ 2008, 
\href{http://dx.doi.org/10.1086/588755}{\JournalTitle{ApJL}, 178, 137 }

\bibitem[Sun et al. (2014)]{Sun14} Sun X.~H., Gaensler B.~M., Carretti E., et al.,\ 2014, 
\href{http://dx.doi.org/10.1093/mnras/stt2110} {\JournalTitle{MNRAS}, 437, 2936}

\bibitem[Yan \& Lazarian(2002)]{Yan02} Yan, H., \& Lazarian, A.,\ 2002, 
\href{http://adsabs.harvard.edu/abs/2002PhRvL..89B1102Y}{\JournalTitle{PhRvL}, 89, 281102} 

\bibitem[Yuen \& Lazarian (2017)]{YL17} Yuen K.~H., \& Lazarian A.,\ 2017,   
 \href{http://dx.doi.org/10.3847/2041-8213/aa6255}{\JournalTitle{ApJL}, 837, L24} 
 
 \bibitem[Yuen et al. (2018) ]{Yuen18} Yuen K.~H., Chen J., Hu Y., et al.,\ 2018, 
  \href{http://adsabs.harvard.edu/abs/2018ApJ...865...54Y}{\JournalTitle{ApJ}, 865, 54}
 
 \bibitem[Van Eck et al. (2017)]{VanEck17} Van Eck C.~L., Haverkorn M., Alves M.~I.~R., et al.,\ 2017, 
 \href{http://dx.doi.org/10.1051/0004-6361/201629707}{\JournalTitle{A\&A}, 597, A98}
 
 \bibitem[Waelkens, Schekochihin \& En{\ss}lin (2009)]{Waelkens09} Waelkens A.~H., Schekochihin A.~A., \& En{\ss}lin, T.~A.,\ 2009,
\href{http://adsabs.harvard.edu/abs/2009MNRAS.398.1970W} {\JournalTitle{MNRAS}, 398, 1970} 

\bibitem[Xu \& Zhang (2016)]{XuZhang16} Xu S., \& Zhang B.,\ 2016, 
\href{http://dx.doi.org/10.3847/0004-637X/824/2/113}{\JournalTitle{ApJ}, 824, 113} 
 
\bibitem[Zhang, Lazarian \& Xiang (2018)]{ZhangL18} Zhang J.-F., Lazarian A., \& Xiang F.-Y.,\ 2018,
\href{http://dx.doi.org/10.3847/1538-4357/aad182}{\JournalTitle{ApJ}, 863, 197 }

\bibitem[Zhang et al. (2016)]{Zhang16} Zhang J.-F., Lazarian A., Lee H., \& Cho J.,\ 2016, 
\href{http://dx.doi.org/10.3847/0004-637X/825/2/154}{\JournalTitle{ApJ}, 825, 154}



\end{thebibliography}
\end{document}